\documentclass[useAMS,usenatbib]{mn2e}

\usepackage{graphicx}
\usepackage{amsmath}
\usepackage{amssymb}
\usepackage{color}
\usepackage{subfig}

\newcommand\msun{\, \rm M_\odot}
\newcommand\rsun{\, \rm R_\odot}
\newcommand\kms{\, \rm km\,s^{-1}}

\newcommand\kpc{{\, \rm kpc}}
\newcommand\yr{{\, \rm yr}}
\newcommand\gyr{{\, \rm Gyr}}
\newcommand\au{{\, \rm AU}}

\newcommand\rmin{{r_{\rm min}}}
\newcommand\aout{{a_{\rm out}}}
\newcommand\ain{{a_{\rm in}}}

\title[Tidal breakup of quadruple stars in the GC]{Tidal breakup of quadruple stars in the Galactic Centre}
\author[G. Fragione]{Giacomo Fragione$^{1}$\thanks{E-mail: giacomo.fragione@mail.huji.ac.il}\\
$^{1}$Racah Institute for Physics, The Hebrew University, Jerusalem 91904, Israel}

\begin{document}

\maketitle

\begin{abstract}
The most likely origin of hypervelocity stars (HVSs) is the tidal disruption of a binary star by the supermassive black hole (MBH) in the Galactic Centre (GC). However, HE0437-5439, a $9\msun$ B-type main-sequence star moving with a heliocentric radial velocity of about $720\kms$ at a distance of $\sim60\kpc$, and the recent discovered hypervelocity binary candidate (HVB), traveling at $\sim 570\kms$, challenge this standard scenario. Recently, Fragione \& Gualandris (2018) have demonstrated that the tidal breakup of a triple star leads to an insufficient rate. Observations show that quadruple stars made up of two binaries orbiting their common center of mass (the so-called 2+2 quadruples) are $\approx 4\%$ of the stars in the solar neighborhood. Although rarer than triples, 2+2 quadruple stars may have a role in ejecting HVBs as due to their larger energy reservoir. We present a numerical study of 2+2 quadruple disruptions by the MBH in the GC and find that the production of HVBs has a probability $\lesssim 2-4\%$, which translates into an ejection rate of $\lesssim 1\gyr^{-1}$, comparable to the triple disruption scenario. Given the low ejection rate, we suggest that alternative mechanisms are responsible for the origin of HVBs, as the ejection from the interaction of a young star cluster with the MBH in the GC and the origin in the Large Magellanic Cloud.
\end{abstract}

\begin{keywords}
Galaxy: centre -- Galaxy: kinematics and dynamics -- stars: kinematics and dynamics -- galaxies: star clusters: general
\end{keywords}

\section{Introduction}

The existence of hypervelocity stars (HVSs) was predicted for the first time by \citet{hills88} as the consequence of the tidal separation of a binary stars approaching too close a massive black hole (MBH). Nearly a couple of decades after, \citet{brw05} observed the first HVS in the Galactic halo, moving with a heliocentric radial velocity of $\sim 700\kms$. Only $\approx 20$ HVSs have been observed mainly by the Multiple Mirror Telescope survey \citep{brw06,brw14} and the LAMOST survey \citep{zhc14}, while a larger sample is expected from the astrometric European satellite \textit{Gaia} \citep{brw15,mar17}. Although rare, HVSs can provide useful information both about their formation environment and the Galactic mass distribution \citep{yut03,gnedin2010,frl17,fgg17}.

Alternative mechanisms have been proposed to explain the observed population of HVSs, as encounters with a massive black hole binary in the GC \citep{yut03} and the tidal interaction of stars clusters with a single or binary MBHs \citep*{cap15,fra16,fck17}. Moreover, a few HVSs are outliers in the overall standard Hills picture, as HE0437-5439, a $9\msun$ B-type main-sequence whose travel time is longer than its main sequence lifetime \citep{ede05}. \citet{per09} suggested that HE0437-5439 can come from the GC if the star was originally a hypervelocity binary (HVB), as a consequence of the tidal breakup of a triple star, that later evolved into a rejuvenated blue straggler. \citet{GPZ2007} proposed that such a star may have been produced in the Large Magellanic Cloud via the standard Hills scenario. \citet{fck17} suggested the tidal interaction of a compact young star cluster infalling onto the MBH as a possible origin. Recently, \citet{erkd18} have used proper motion measurements from \textit{Gaia} DR2 and have suggested that HE0437-5439 is likely to come from the centre of the Large Magellanic Cloud, which should host a black hole of thousands solar masses. The recent discovery of a candidate HVB by \citet{nem16} put attention back on all these alternative scenarios. If not mediated by clusters, the ejection of an HVB from the GC would require the breakup of a triple star, whose rate is $\lesssim 1$ Gyr$^{-1}$, making unlikely a triple disruption origin \citep{fgu18}. 

\begin{table*}
\caption{Models: name, mass of the stars of the first binary ($m_1$-$m_2$), mass of the stars of the second binary ($m_3$-$m_4$), star radius ($R_*$), first binary semi-major axis ($a_{in,1}$), second binary semi-major axis ($a_{in,2}$), outer quadruple semi-major axis ($\aout$).}
\centering
\begin{tabular}{|l|c|c|c|c|c|c|}
\hline
Name & $m_1$-$m_2$ ($\msun$) & $m_3$-$m_4$ ($\msun$) & $R_*$ & $a_{in,1}$ (AU) & $a_{in,2}$ (AU) & $\aout$ (AU)\\
\hline
Model 1			& $3$ 		& $3$		& $0$ & $0.05$-$0.1$	& $0.05$-$0.1$	& $0.5$ \\
Model 1r		& $3$ 		& $3$		& yes & $0.05$-$0.1$	& $0.05$-$0.1$	& $0.5$ \\
Model 2			& $3$ 		& $3$		& $0$ & $0.05$-$0.1$	& $0.05$		& $0.5$\\
Model 2r		& $3$ 		& $3$		& yes & $0.05$-$0.1$	& $0.05$		& $0.5$\\
Model 3			& $3$ 		& $3$		& $0$ & $0.05$			& $0.05$		& $0.5$-$1.0$\\
Model 3r		& $3$ 		& $3$		& yes & $0.05$			& $0.05$		& $0.5$-$1.0$\\
Model 4			& $1$-$4$	& $1$-$4$	& $0$ & $0.05$			& $0.05$ 		& $0.5$\\
Model 4r		& $1$-$4$	& $1$-$4$	& yes & $0.05$			& $0.05$ 		& $0.5$\\
Model 5			& $3$ 		& $3$		& $0$ & $0.025$-$0.05$	& $0.025$-$0.5$	& $0.25$ \\
Model 6			& $3$ 		& $3$		& $0$ & $0.025$-$0.05$	& $0.025$		& $0.25$\\
Model 7			& $3$ 		& $3$		& $0$ & $0.025$			& $0.025$		& $0.25$-$0.5$\\
Model 8			& $1$-$4$	& $1$-$4$	& $0$ & $0.025$			& $0.025$ 		& $0.25$\\
\hline
\end{tabular}
\label{tab:model}
\end{table*}

In this paper, we study the ejection of HVBs as a consequence of close encounters between a quadruple 2+2 star and the MBH in the GC by means of high-precision scattering experiments. Observations have shown that more than $50$\% of stars have at least one stellar companion \citep{duc13,tok14a,tok14b}, while \citet{rid15} found a relatively large abundance ($\approx 4\%$) of $2$+$2$ quadruples. The dynamics of 2+2 quadruple systems has been under scrutiny to explain several astrophysical phenomena. \citet{tok18} analysed some 2+2 systems and found that they may be formed either sequentially by disk fragmentation with subsequent accretion and migration or by a cascade hierarchical fragmentation of a rotating cloud. \citet{fan17} studied 2+2 quadruple dynamics in the contest of white dwarf-white dwarf mergers and collisions connected to Type Ia supernovae. Finally, \citet{set18} investigated the orbital synchronization of 2+2 quadruples and their implications for gravitational waves. Given the dynamical relevance of 2+2 quadruple stars and their observed high frequency, quadruple disruptions in the GC should not be rare.

The paper is organised as follows. In Section 2, we describe the methods and initial conditions we used in our $N$-body integrations. In Section 3, we present the results of the scattering experiments. Finally, we discuss the implications of our findings and present our conclusions in Section 4.

\section{Method}
\label{sect:meth}

In the standard \citet{hills88} scenario, a binary of mass $m$ and semi-major axis $a$ is tidally disrupted whenever approaches the MBH of mass $M$ within a distance of the order of the binary tidal radius
\begin{equation}
r_t\approx \left(\frac{M}{m}\right)^{1/3} a\ .
\label{eqn:rt}
\end{equation}
A binary, starting from an initial distance $D$ with a transverse speed $v$, will approach the MBH to a minimum distance $\rmin$ given by momentum conservation
\begin{equation}
v\,D=\left(\frac{GM}{\rmin}\right)^{1/2} \rmin\ .
\label{eqn:rmin}
\end{equation}
In the case $\rmin \lesssim r_t$, the binary is disrupted and the former companions separated. As discussed in \citet{fgu18}, there are three possible outcomes for the tidal breakup of a binary (production of an HVS and a S-star, production of 2 S-stars, capture of the whole binary), while in the case of a triple star even seven outcomes are possible. We define as S-star the single star or binary star that remains bound to the MBH. In the case of a 2+2 quadruple star-MBH encounter the number of possible channels is larger. We identify $5$ main outcome classes based on the multiplicity of stars after the close interaction with the MBH:
\begin{itemize}
\item 4 single stars (4S);
\item 1 binary star and 2 single stars (1B-2S);
\item 2 binary stars (2B);
\item 1 triple star and 1 single star (1T-1S);
\item 1 quadruple star (1Q).
\end{itemize}
Each class of outcomes has several subclasses of outcomes, according to the number of objects that remains bound or becomes unbound to the MBH. For instance, in the second channel, the final multiplicity provides 1 binary star and 2 single stars and contains 6 suboutcomes: - all the objects are bound to the MBH (two S-stars and one binary S-star); - all the objects are unbound to the MBH (two HVSs and one HVB); - the single stars are bound to the MBH and the binary star is unbound (two S-stars and one HVB); - the single stars are unbound to the MBH and the binary is bound (two HVSs and one binary S-star); - one single star is bound to the MBH and the second single star and the binary star are unbound (one HVS, one single S-star and one HVB); - one single star is unbound to the MBH and the second single star and the binary star are bound (one HVS, one S-star and one binary S-star).

We integrate the system for a total time $T=D/v$ to resolve all possible channels for all the scattering events. When the scattering event is completed, we use energy considerations to distinguish among the possible outcomes, as follows. We first find the most bound pair among the $4$ stars by computing pairwise relative energies. If no pair of stars is bound, we conclude that the scattering outcome is the channel 4S. If at least one bound pair is found, we check if the last two stars form a second bound pair. If this is the case, we classify the scattering outcome as 4Q, if the two binaries are bound, or as 2B, if they are unbound; otherwise, we determine if any of the remaining two stars forms a bound triple system with the most bound binary star. If it does, we conclude we are in the channel 1T-1S. If no triple configuration is found, we classify the outcome as 1B-2S. After the first classification based on the multiplicity of the outcomes, we compute the relative energy of each object with respect to the MBH. If positive, they are classified as unbound hypervelocity objects, otherwise they are classified as bound S-objects.

\begin{figure} 
\centering
\includegraphics[scale=0.525]{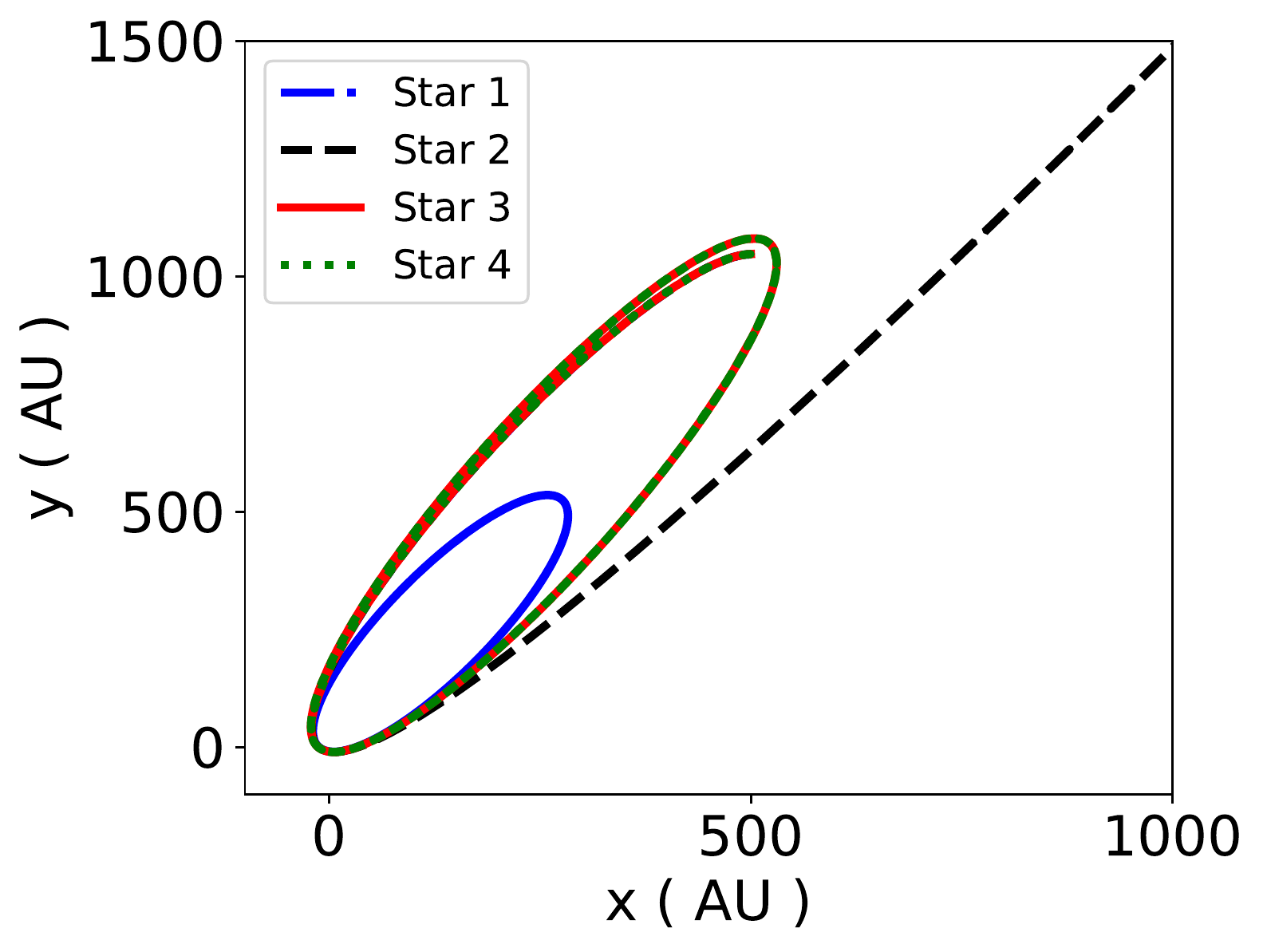}
\caption{Example of scattering for Model 1 in the case $a_{in,1}=a_{in,2}=0.05$ AU. The initial distance on the x-axis is $500$ AU, with the MBH at the origin of the reference frame. The quadruple star gets tidally disrupted by the MBH, leaving a single S-star and a binary S-star on high eccentricity orbits, and a HVS.}
\label{fig:scattex}
\end{figure}

We initialise the initial conditions of the centre of mass of the 2+2 quadruple following the prescriptions of \citet{gl06,gl07}. Each quadruple star starts from a distance $D=10^3\times \aout$ (with respect to the MBH) with initial transverse velocity $v = 250\kms$ \citep{hills88}. Then, we generate the maximum initial impact parameter $b_{max}$ such that the pericentre of the quadruple is $\lesssim r_t$ (Eq.\,\ref{eqn:rmin}). To take into account the gravitational focusing \citep{brm06}, we randomly sample from a probability distribution $f(b)\propto b$ in the pericentre distance.

The initial conditions for the numerical experiments have been set as follows (see also Table\,\ref{tab:model}):
\begin{itemize}
\item The mass of the MBH is fixed to $M=4\times 10^6\msun$ \citep{gil09}.
\item Stellar masses of stars in the quadruple are set to $m_*=1$, $2$, $3$, $4 \msun$.
\item Stellar radii are computed from \citep{dem91}
\begin{equation}
R_*=
\begin{cases}
1.06\ (m_*/\msun)^{0.945}\rsun& \text{$ m_*< 1.66\msun$}\ ,\\
1.33\ (m_*/\msun)^{0.555}\rsun& \text{$ m_*> 1.66\msun$}\ .
\end{cases}
\end{equation}
If the finite stellar radii are taken into account (models marked with "r"), we monitor the relative distances of any two stars during all the duration of the encounter and, if any of them becomes smaller than the sum of the stellar radii, the stars are considered merged and removed from the simulation.
\item The semi-major axis ($a_{in,1}$, $a_{in,2}$) of the two inner binary stars are $0.025$-$0.1\au$.
\item The initial eccentricity of the two inner binaries are set to $e_{\rm in,1}=e_{\rm in,2}=0$.
\item The initial phases $\chi_{1,1}$ and $\chi_{1,2}$ of the inner binaries, which determine the initial position of the stars on each binary orbit, are randomly generated.
\item The angles $\theta_{1,1}$, $\phi_{1,1}$, $\psi_{1,1}$ and $\theta_{1,2}$, $\phi_{1,2}$, $\psi_{1,2}$, which determine the orientation of the two inner binaries orbital plane with respect to the orbital plane of the quadruple centre of mass, are randomly generated.
\item The semi-major axis of the outer orbit is $\aout = 0.25$-$1.0\au$. 
\item The initial eccentricity of the outer orbit is set to $e_{\rm out}=0$.
\item The initial phase $\chi_{2}$ of the outer orbit, which determines the initial position of the inner binaries centres of mass, is randomly generated.
\item The angles $\theta_{2}$, $\phi_{2}$, $\psi_{2}$, which determine the orientation of the orbital plane of the outer binary with respect to the orbital plane of the centre of mass of the quadruple, are randomly generated.
\end{itemize}
As discussed in \citet{fgu18}, we note that the initial eccentricity is expected to influence the relative number of mergers, but not the relative probability of different outcomes since the outputs depend mainly on the initial energy of the system. Finally, we also note that our initial configurations satisfy the stability criterion of hierarchical stars \citep{mar01}.

\begin{figure*} 
\centering
\begin{minipage}{20.5cm}
\subfloat{\includegraphics[scale=0.58]{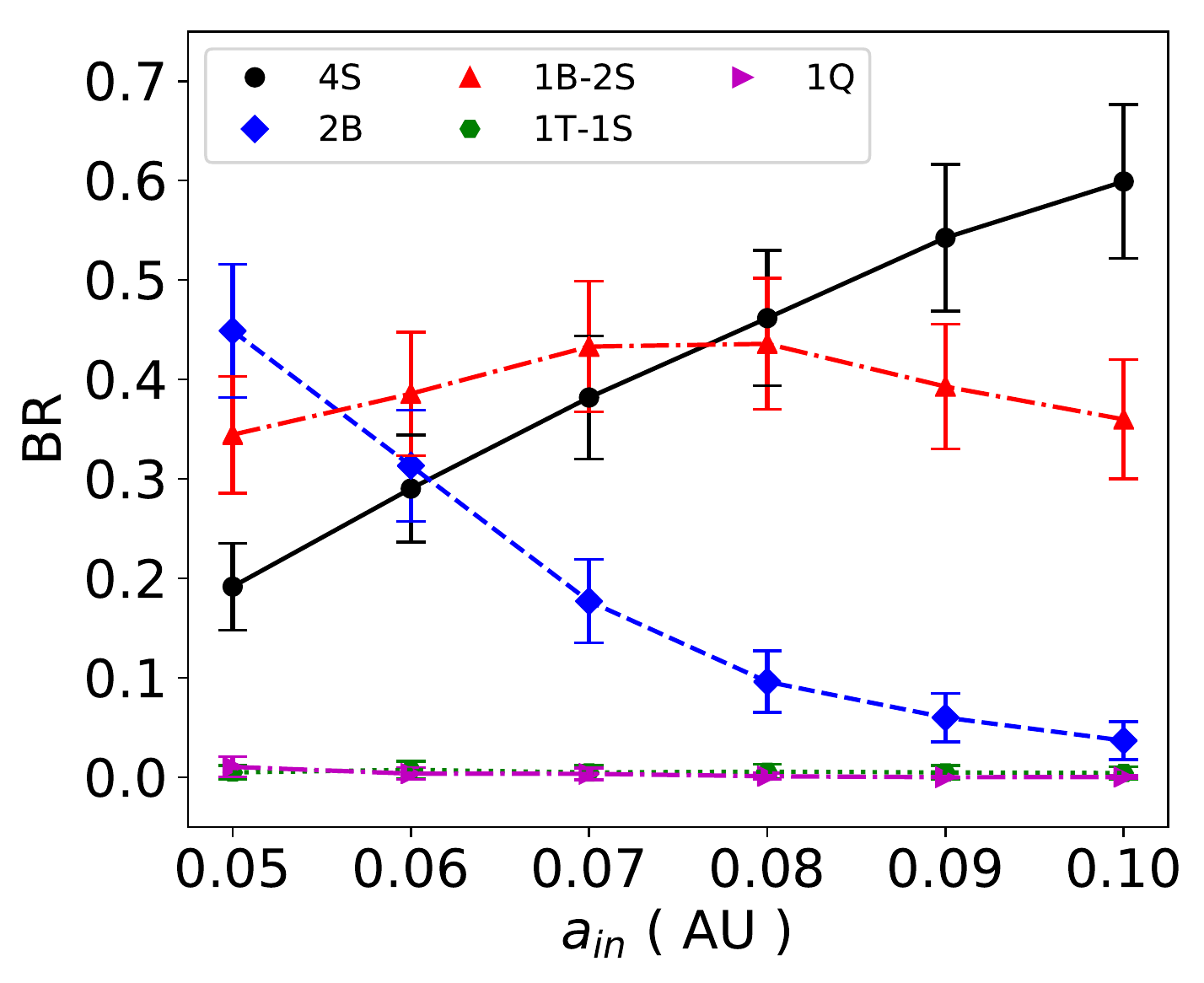}}
\subfloat{\includegraphics[scale=0.58]{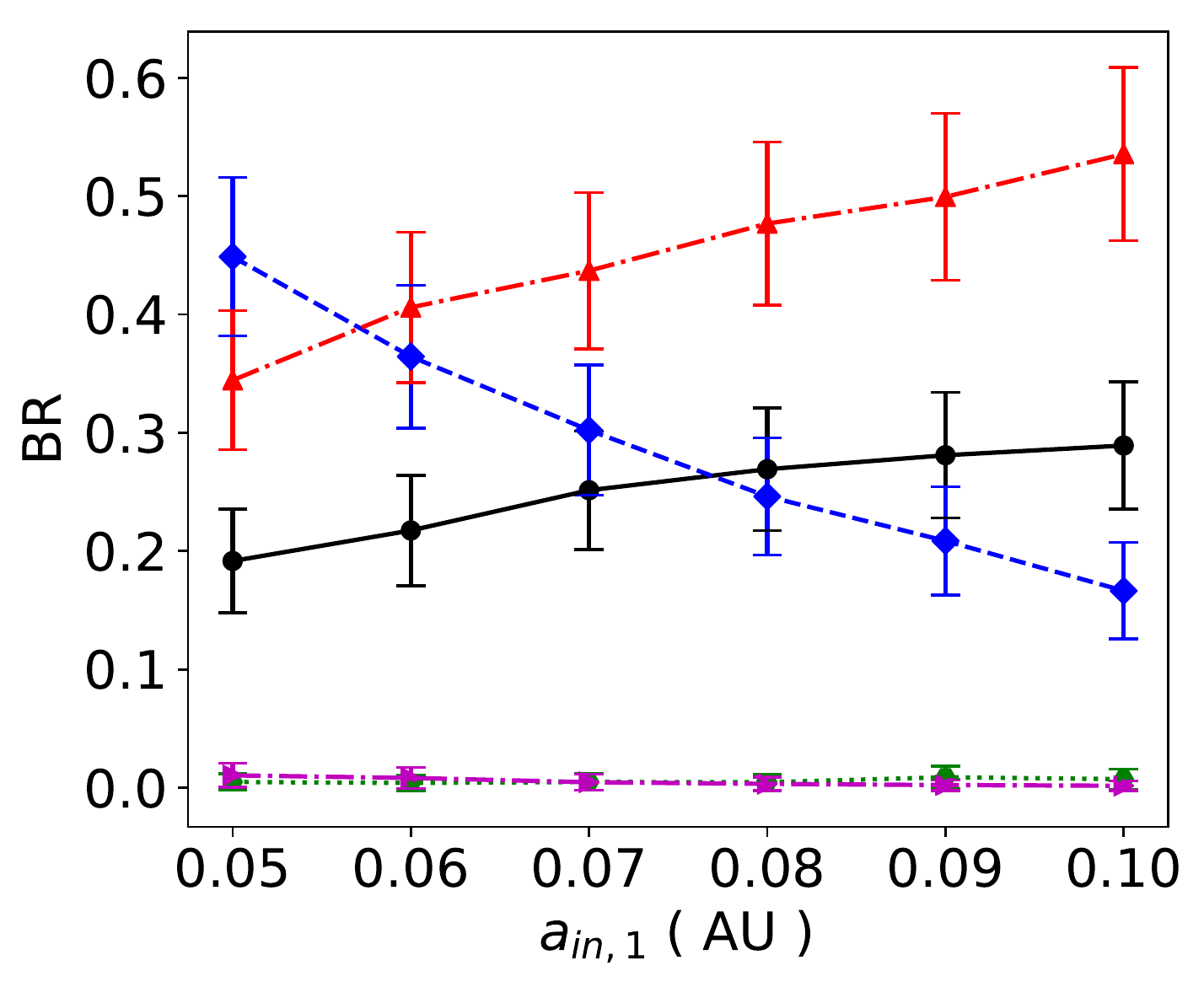}}
\end{minipage}
\begin{minipage}{20.5cm}
\subfloat{\includegraphics[scale=0.58]{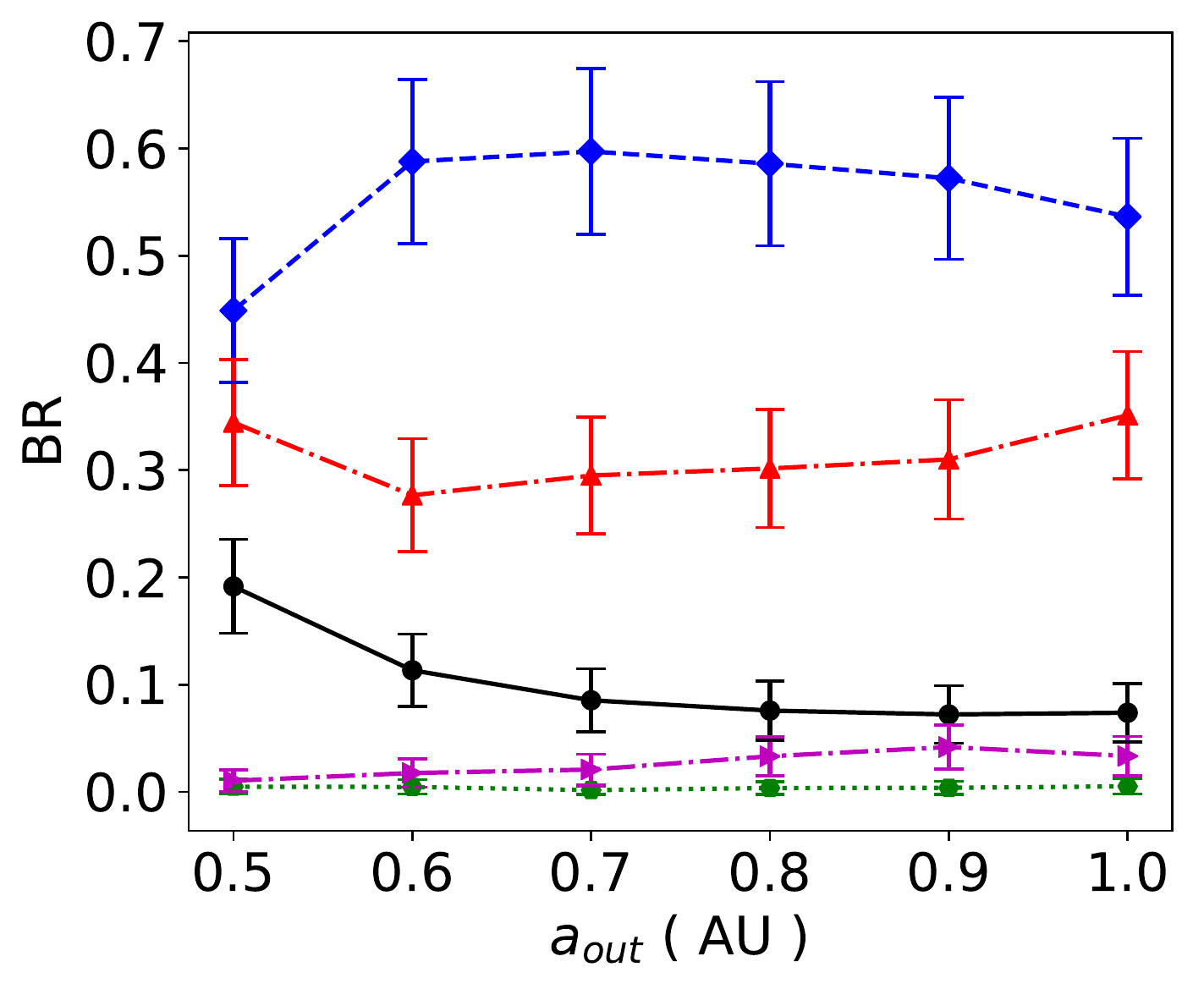}}
\subfloat{\includegraphics[scale=0.58]{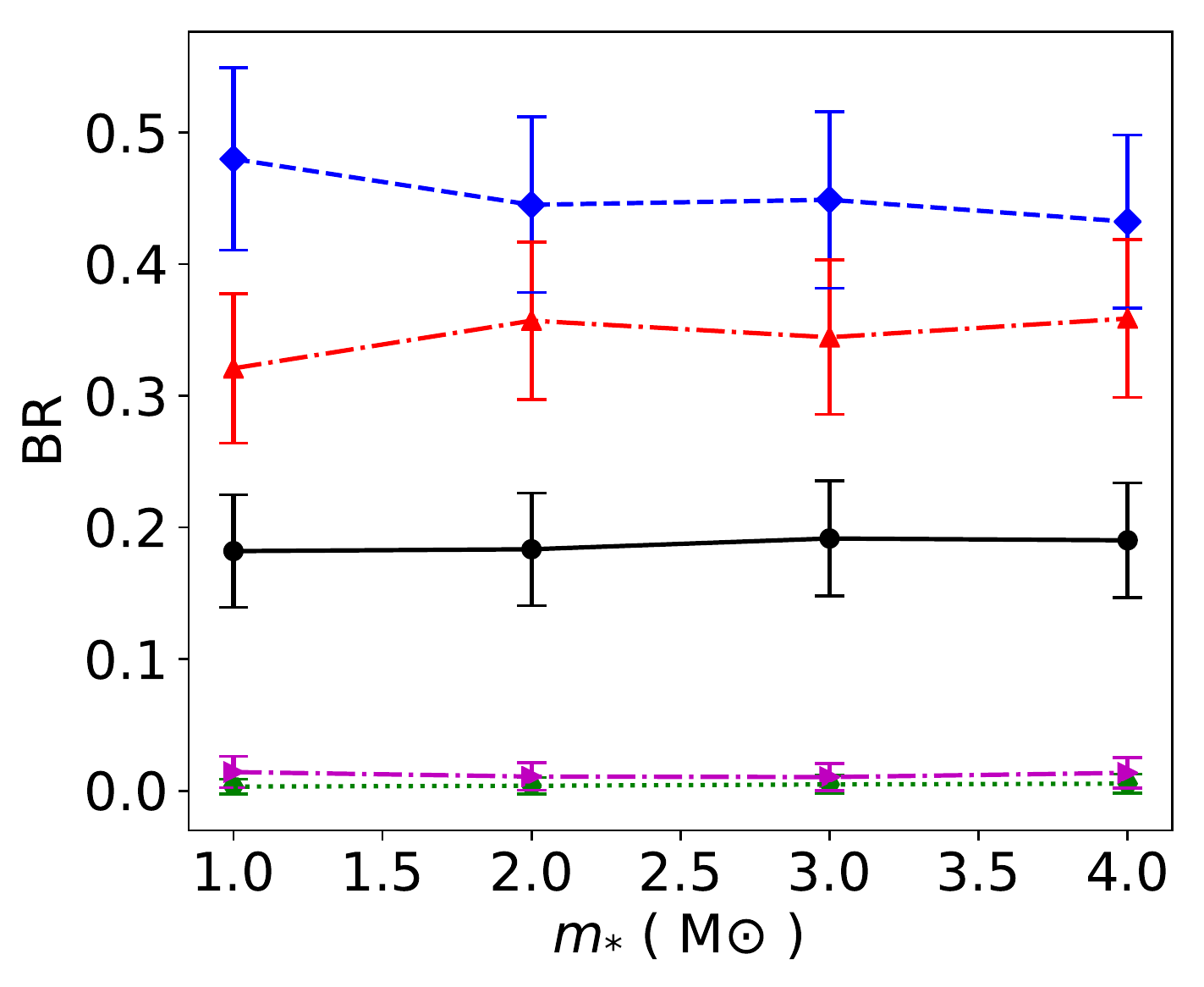}}
\end{minipage}
\caption{Branching ratios for the different channels for Model 1 (top-left) as function of $\ain=a_{in,1}=a_{in,2}$, for Model 2 (top-right) as function of $a_{in,1}$, Model 3 (bottom-left) as function of $\aout$ and Model 4 (bottom-right) as function of $m_*$. Poisson error bars are shown.}
\label{fig:br_all}
\end{figure*}

Given the above set of initial parameters, we integrate the system of the differential equations of motion of the 5-bodies
\begin{equation}
{\ddot{\textbf{r}}}_i=-G\sum\limits_{j\ne i}\frac{m_j(\textbf{r}_i-\textbf{r}_j)}{\left|\textbf{r}_i-\textbf{r}_j\right|^3}\ ,
\end{equation}
with $i=1$,$2$,$3$,$4$,$5$, using the \textsc{archain} code \citep{mik06,mik08}, a fully regularised code able to model the evolution of binaries of arbitrary mass ratios with extreme accuracy. 

\section{Results}

\begin{table*}
\caption{Branching ratios for Model 1 as function of $a_{in}=a_{in,1}=a_{in,2}$. The number indicates the number of objects, the first letter indicates the multiplicity of the object (single, binary, triple or quadruple) and the second letter indicates if the object is bound (S-object) or unbound to the MBH (hypervelocity object). The channels not included in the table have negligible probability.}
\centering
\begin{tabular}{c|c|c|c|c|c|c|c|c|c|c|c|}
\hline
$\ain$ & 4SS & 3SS-1SH & 2SS-2SH & 2BS & 1BS-1BH & 1BS-2SS & 1BS-1SS-1SH & 1BS-2SH & 1BH-2SS & 1TS-1SS & 1QS \\
\hline
0.05 & 0.0642 & 0.0942 & 0.0332 & 0.4384 & 0.0104 & 0.2832 & 0.0464 & 0.002  & 0.0128 & 0.0048 & 0.0104 \\
0.06 & 0.1216 & 0.1426 & 0.0260 & 0.3076 & 0.0056 & 0.3272 & 0.0478 & 0.0002 & 0.0104 & 0.0074 & 0.0036 \\
0.07 & 0.1738 & 0.1854 & 0.0224 & 0.1740 & 0.0030 & 0.3812 & 0.0398 & 0.0008 & 0.0112 & 0.0048 & 0.0034 \\
0.08 & 0.2376 & 0.2066 & 0.0174 & 0.0936 & 0.0024 & 0.3844 & 0.0402 & 0.0002 & 0.0108 & 0.0056 & 0.0010 \\
0.09 & 0.3026 & 0.2276 & 0.0120 & 0.0592 & 0.0008 & 0.3488 & 0.0330 & 0.0002 & 0.0108 & 0.0048 & 0      \\
0.10 & 0.3386 & 0.2488 & 0.0116 & 0.0364 & 0.0004 & 0.3254 & 0.0256 & 0.0088 & 0.0042 & 0.0002 & 0      \\
\hline
\end{tabular}
\label{tab:br1}
\end{table*}

\begin{figure*} 
\centering
\begin{minipage}{20.5cm}
\subfloat{\includegraphics[scale=0.58]{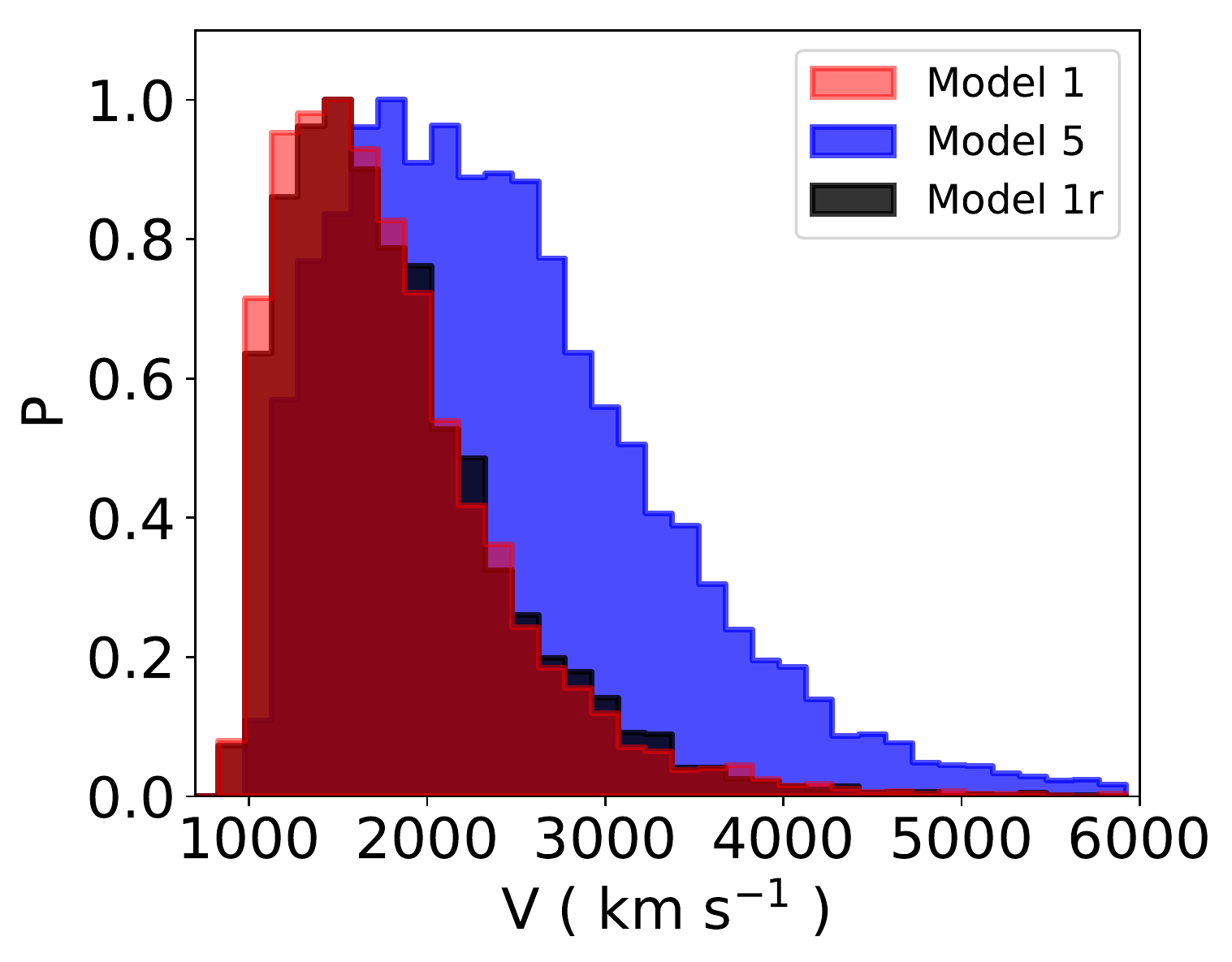}}
\subfloat{\includegraphics[scale=0.58]{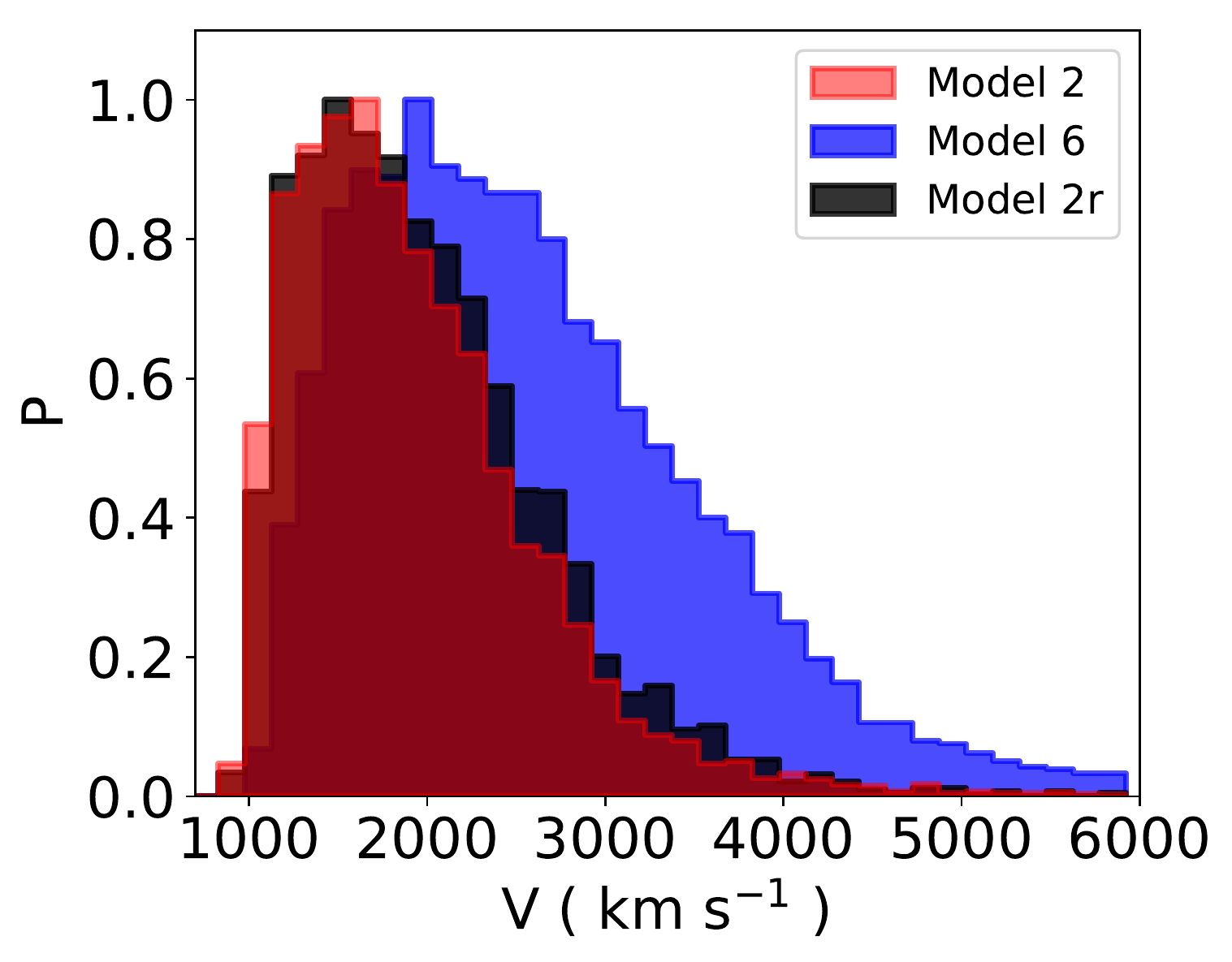}}
\end{minipage}
\begin{minipage}{20.5cm}
\subfloat{\includegraphics[scale=0.58]{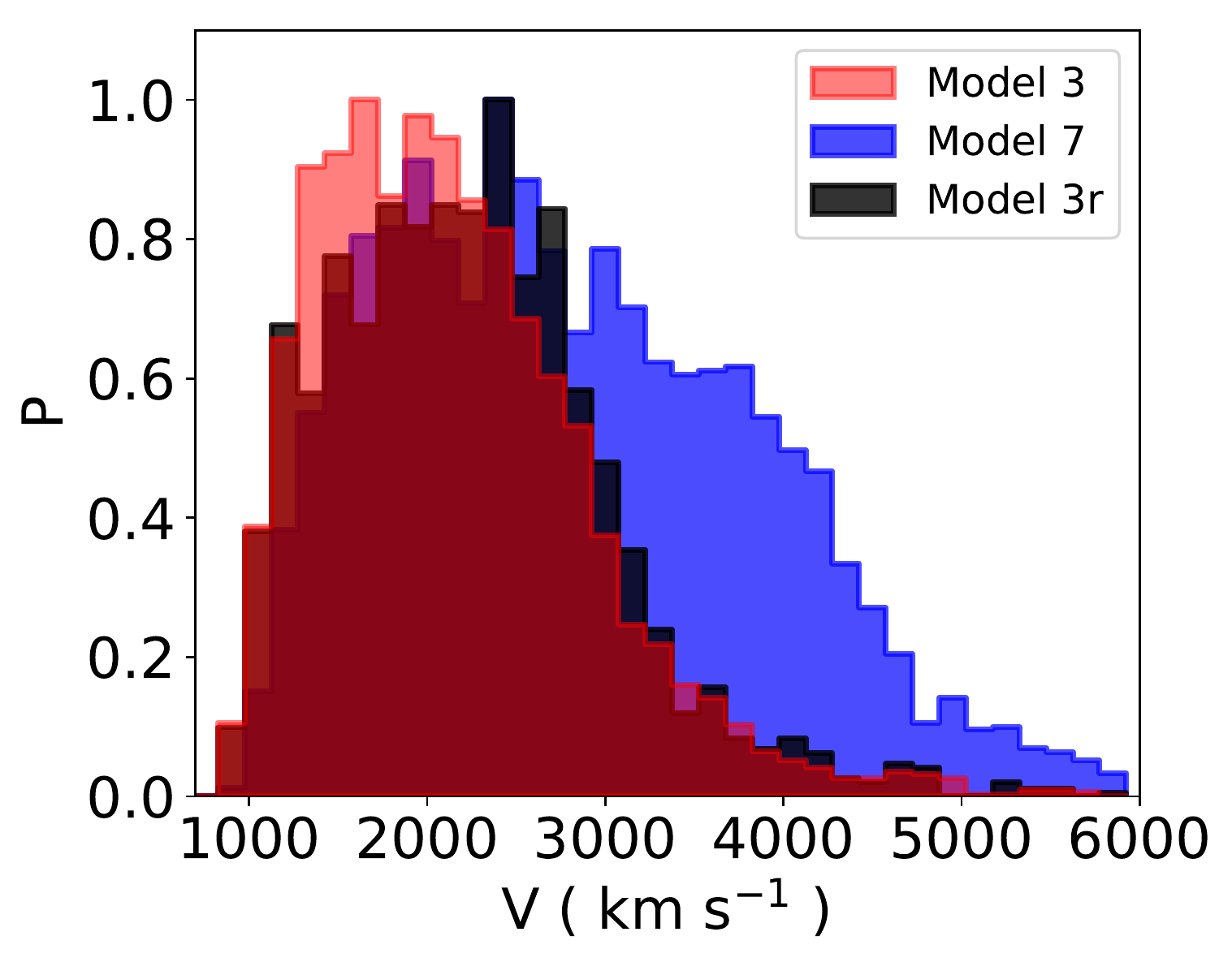}}
\subfloat{\includegraphics[scale=0.58]{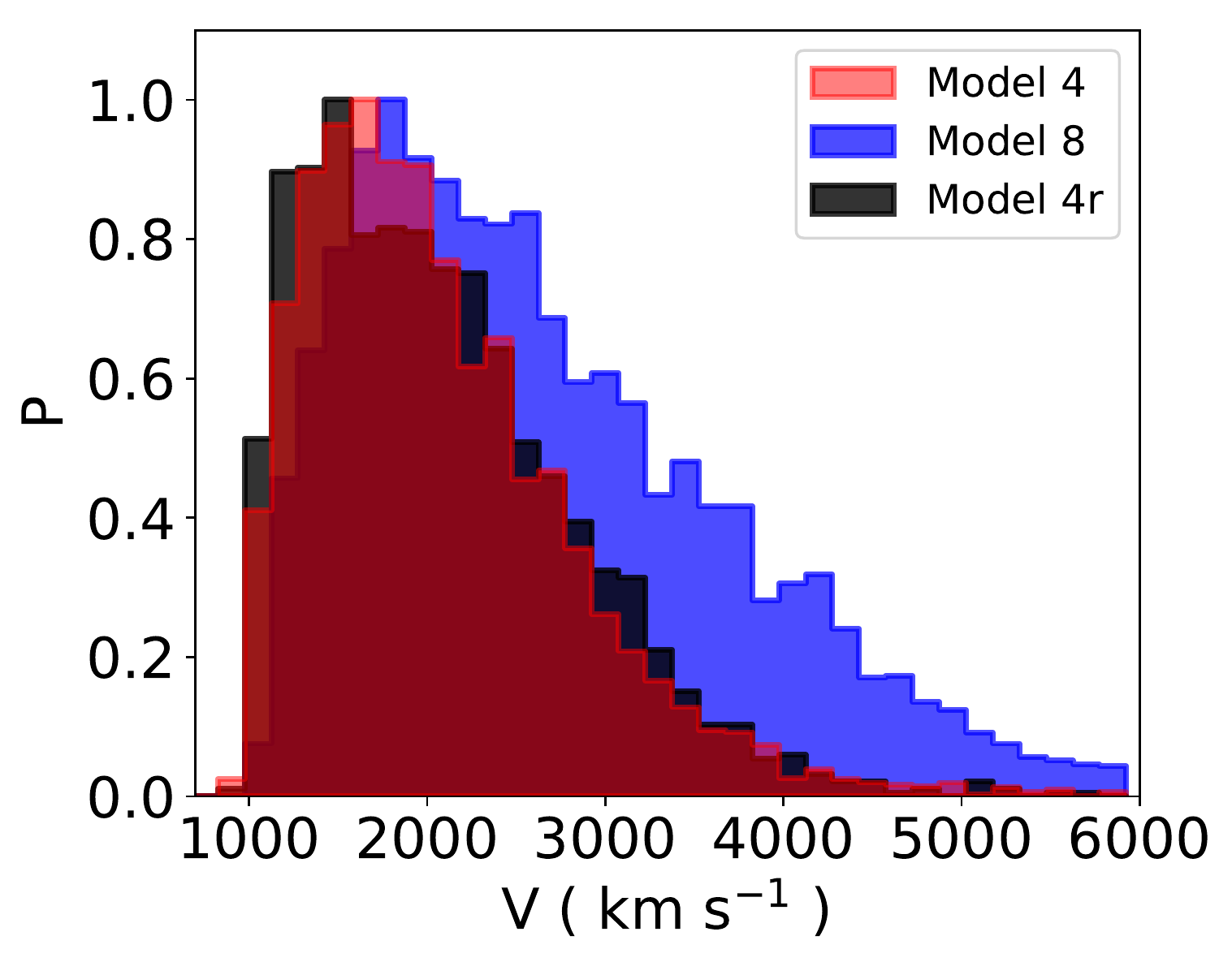}}
\end{minipage}
\caption{Velocity distribution for HVSs for Model 1/1r/5, Model 2/2r/6, Model 3/3r/7 and Model 4/4r/8 when all the $\ain$, $a_{in,1}$, $\aout$ and $m_*$ are considered, respectively.}
\label{fig:vel_sin}
\end{figure*}

\begin{figure*} 
\centering
\begin{minipage}{20.5cm}
\subfloat{\includegraphics[scale=0.58]{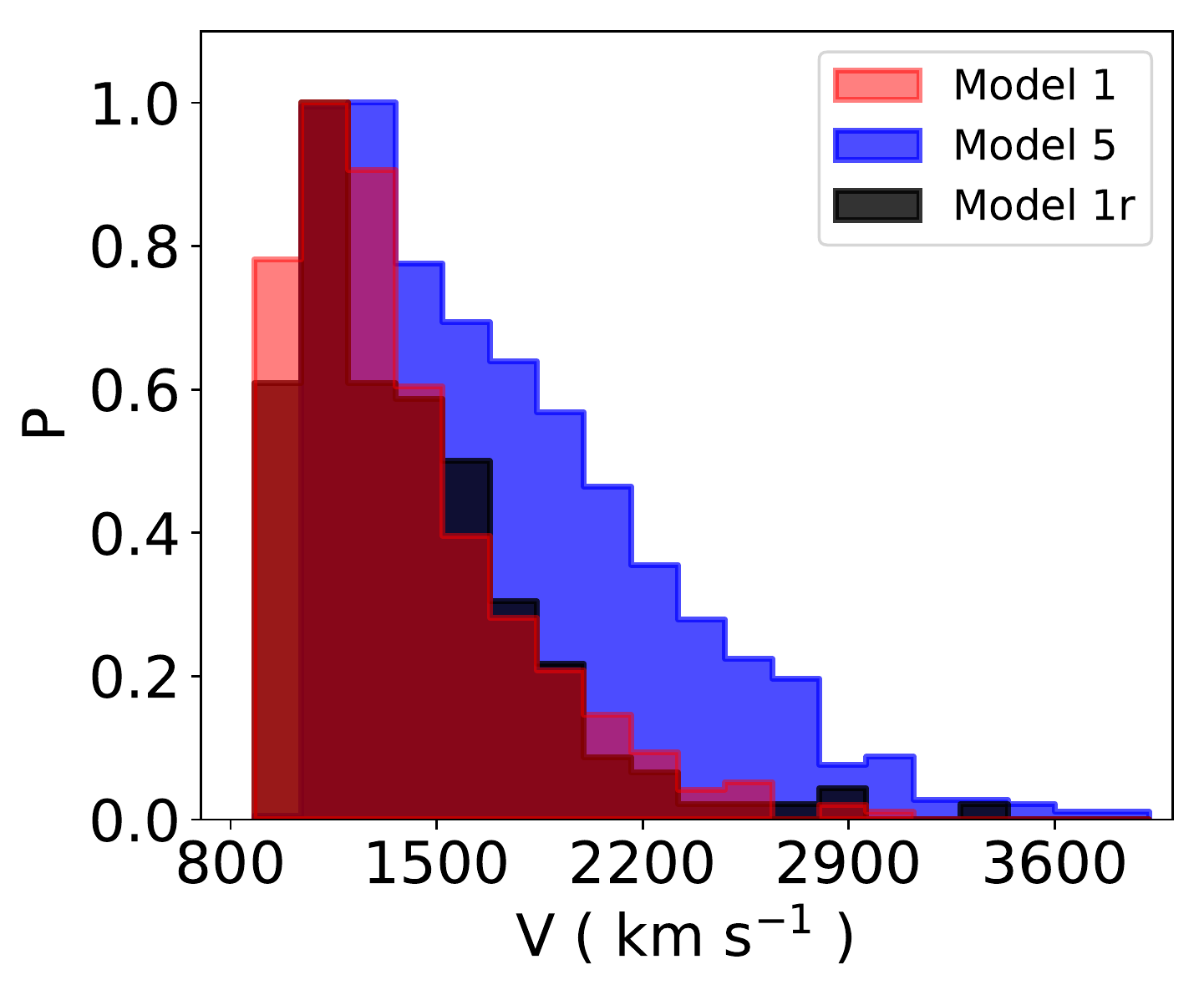}}
\subfloat{\includegraphics[scale=0.58]{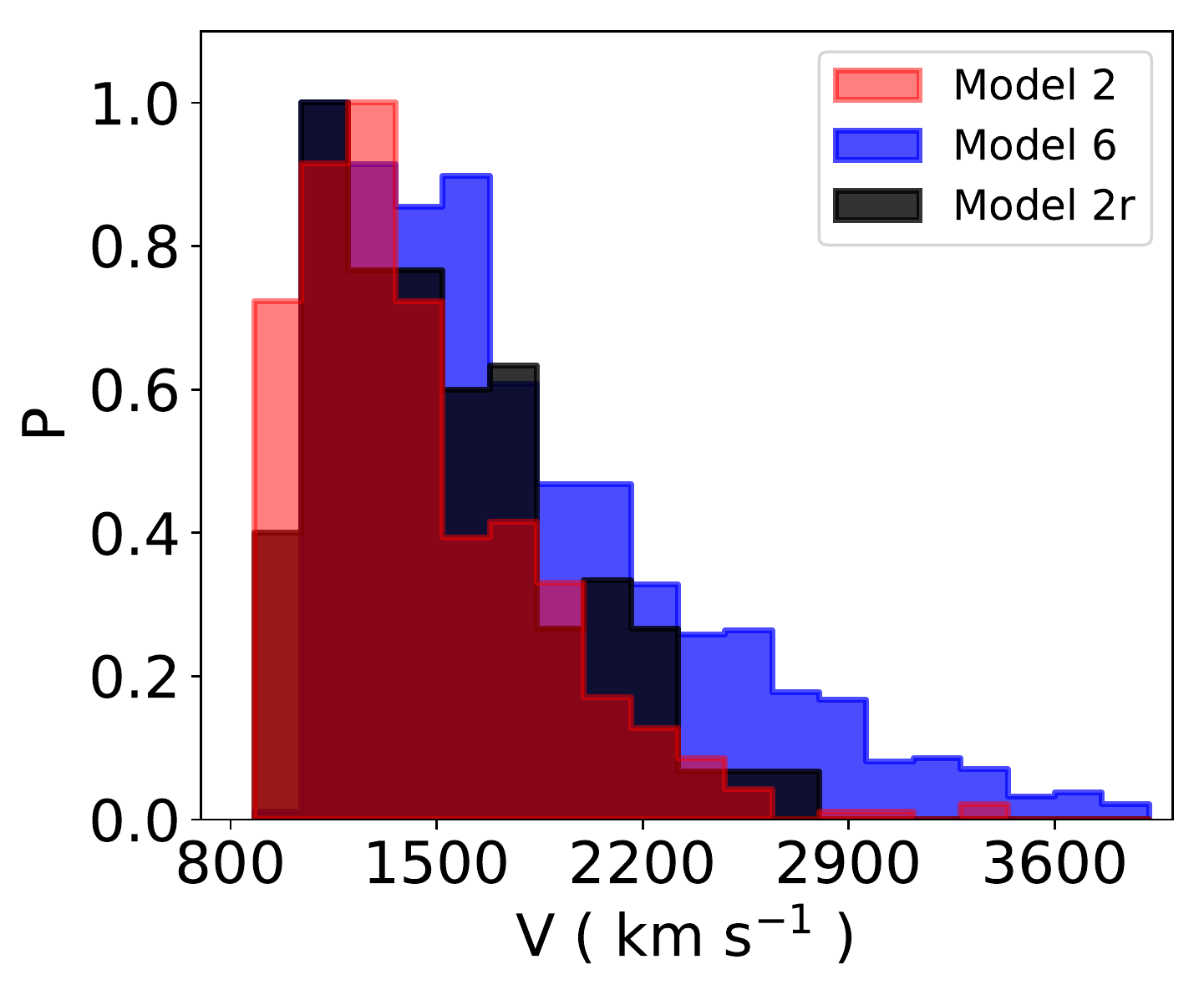}}
\end{minipage}
\begin{minipage}{20.5cm}
\subfloat{\includegraphics[scale=0.58]{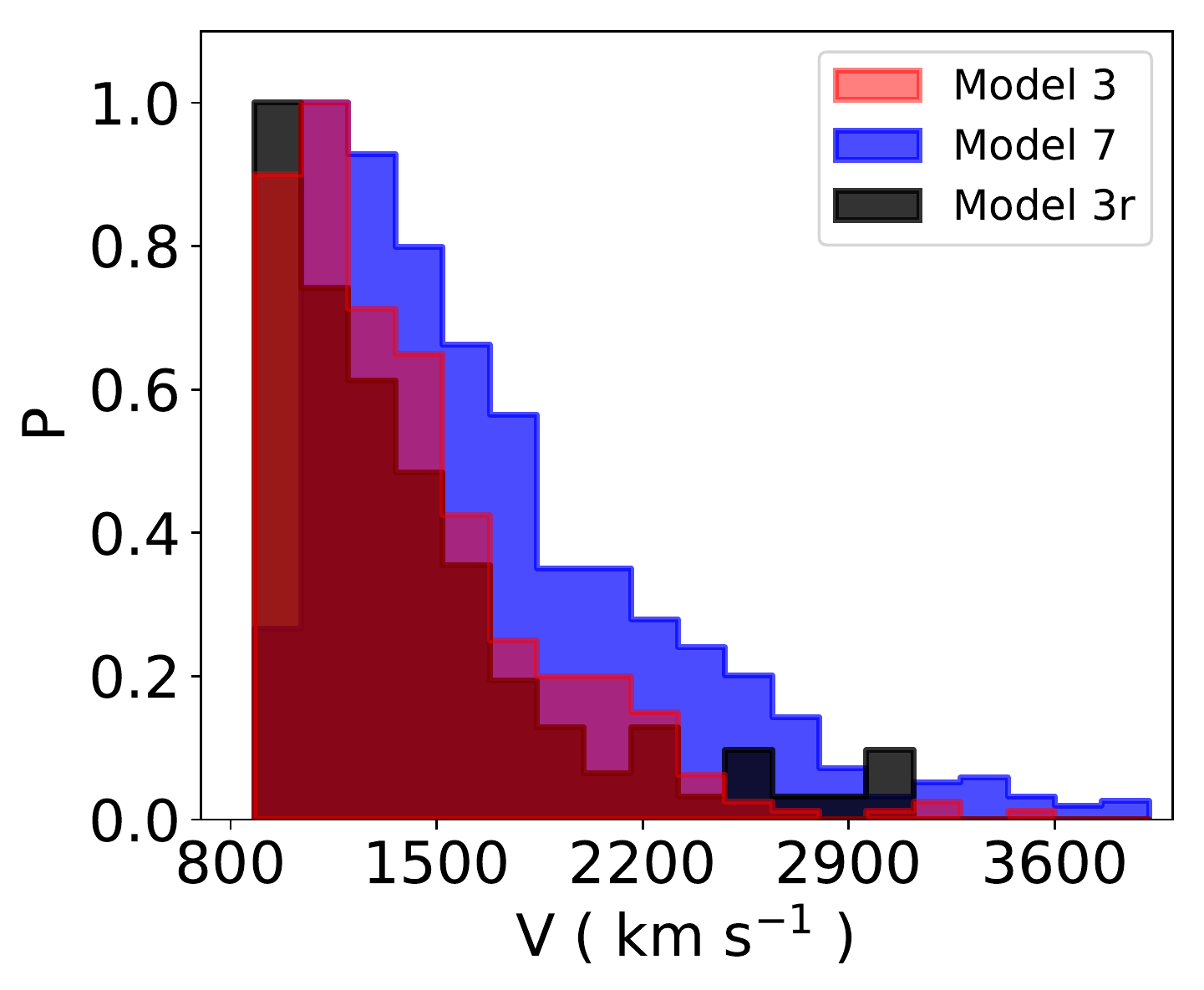}}
\subfloat{\includegraphics[scale=0.58]{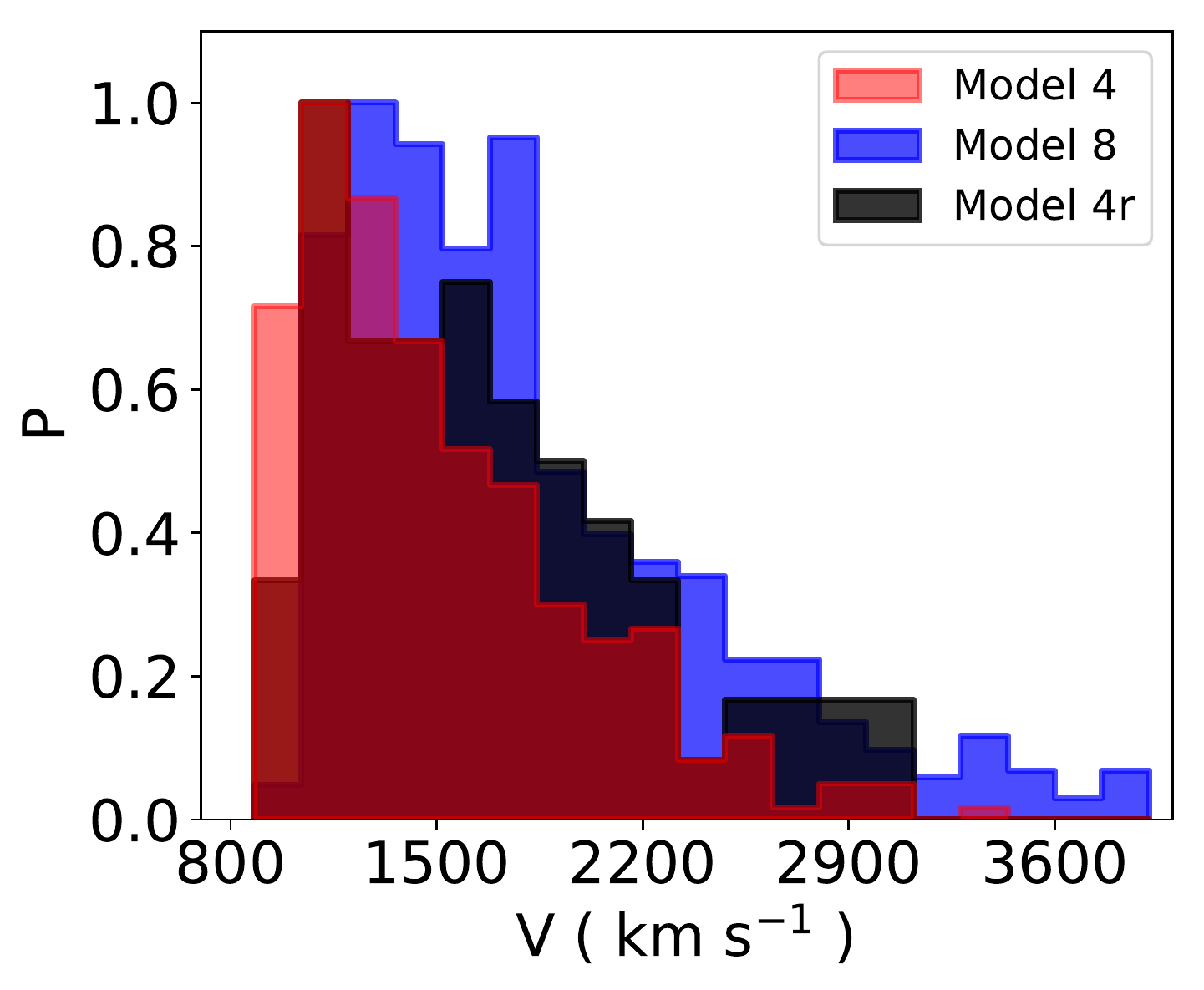}}
\end{minipage}
\caption{Velocity distribution for HVBs for Model 1/1r/5, Model 2/2r/6, Model 3/3r/7 and Model 4/4r/8 when all the $\ain$, $a_{in,1}$, $\aout$ and $m_*$ are considered, respectively.}
\label{fig:vel_bin}
\end{figure*}

\begin{figure*} 
\centering
\begin{minipage}{20.5cm}
\subfloat{\includegraphics[scale=0.58]{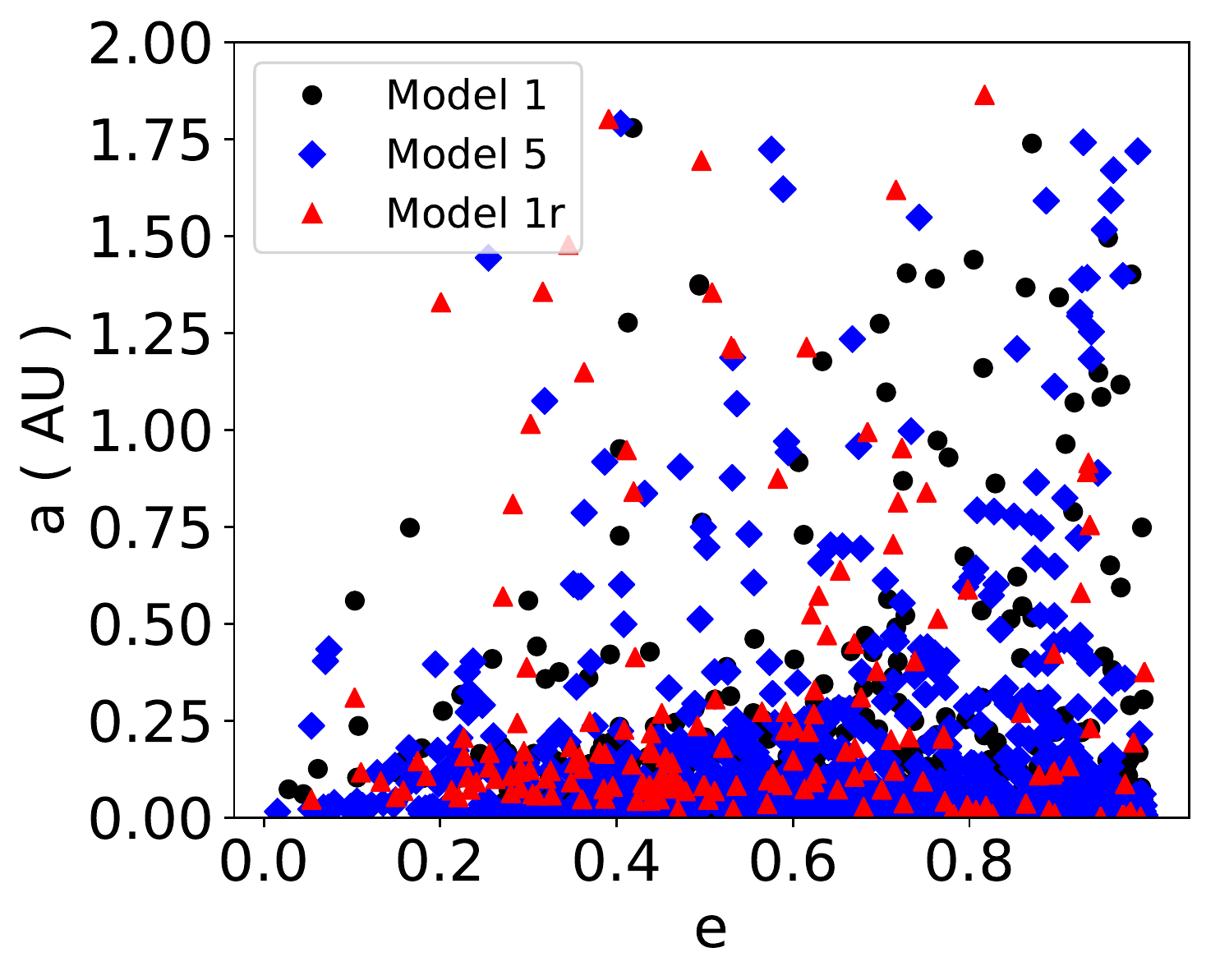}}
\subfloat{\includegraphics[scale=0.58]{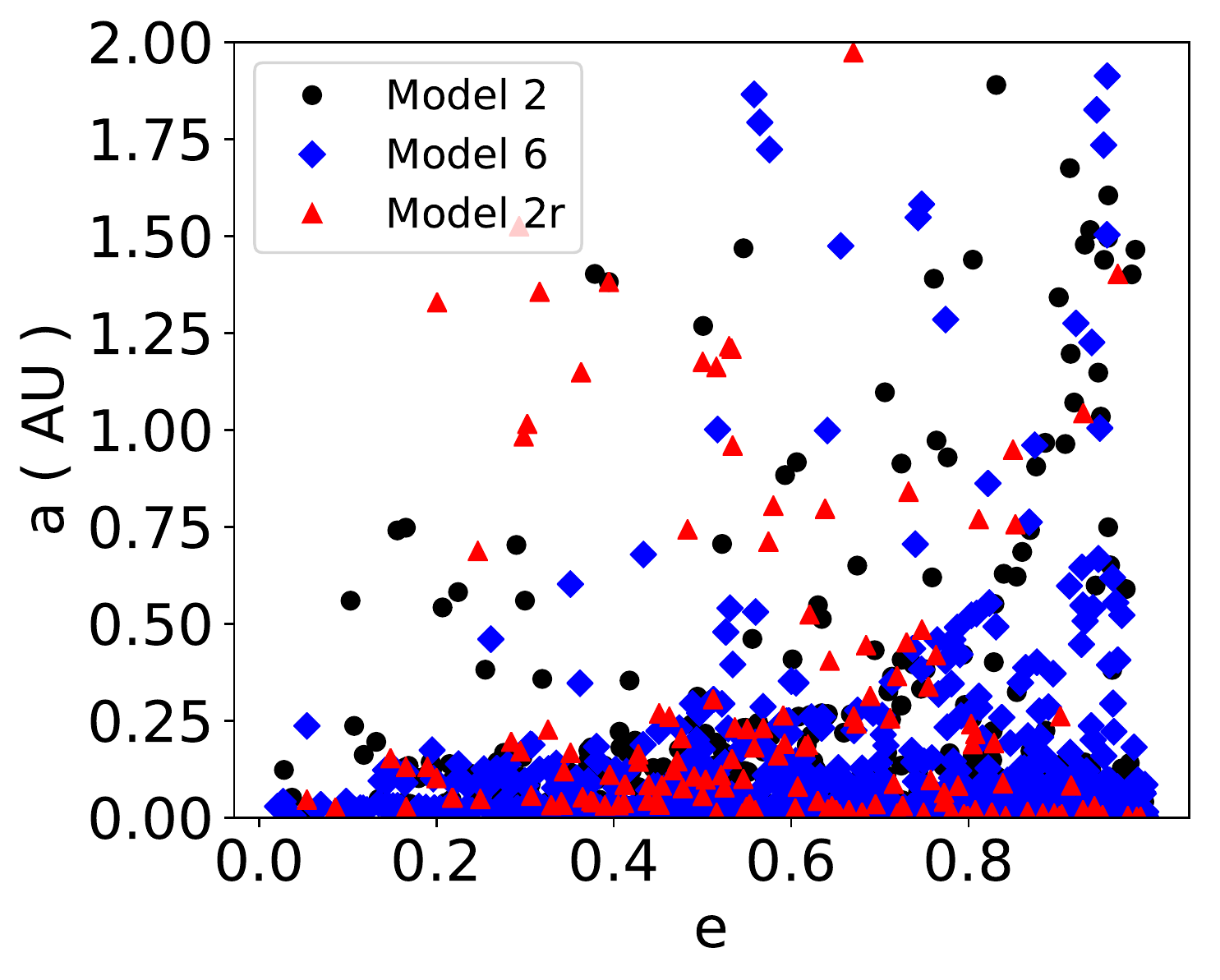}}
\end{minipage}
\begin{minipage}{20.5cm}
\subfloat{\includegraphics[scale=0.58]{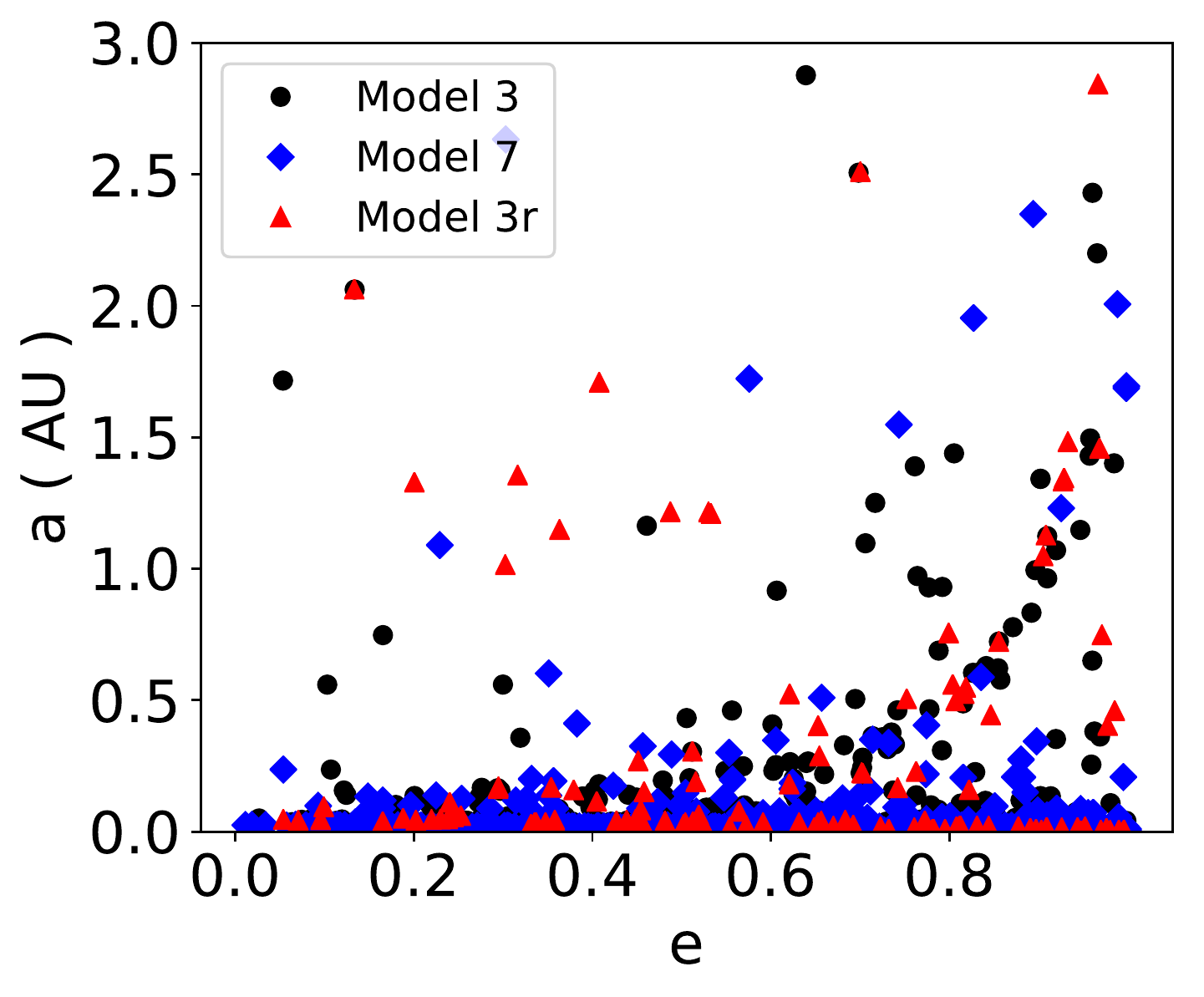}}
\subfloat{\includegraphics[scale=0.58]{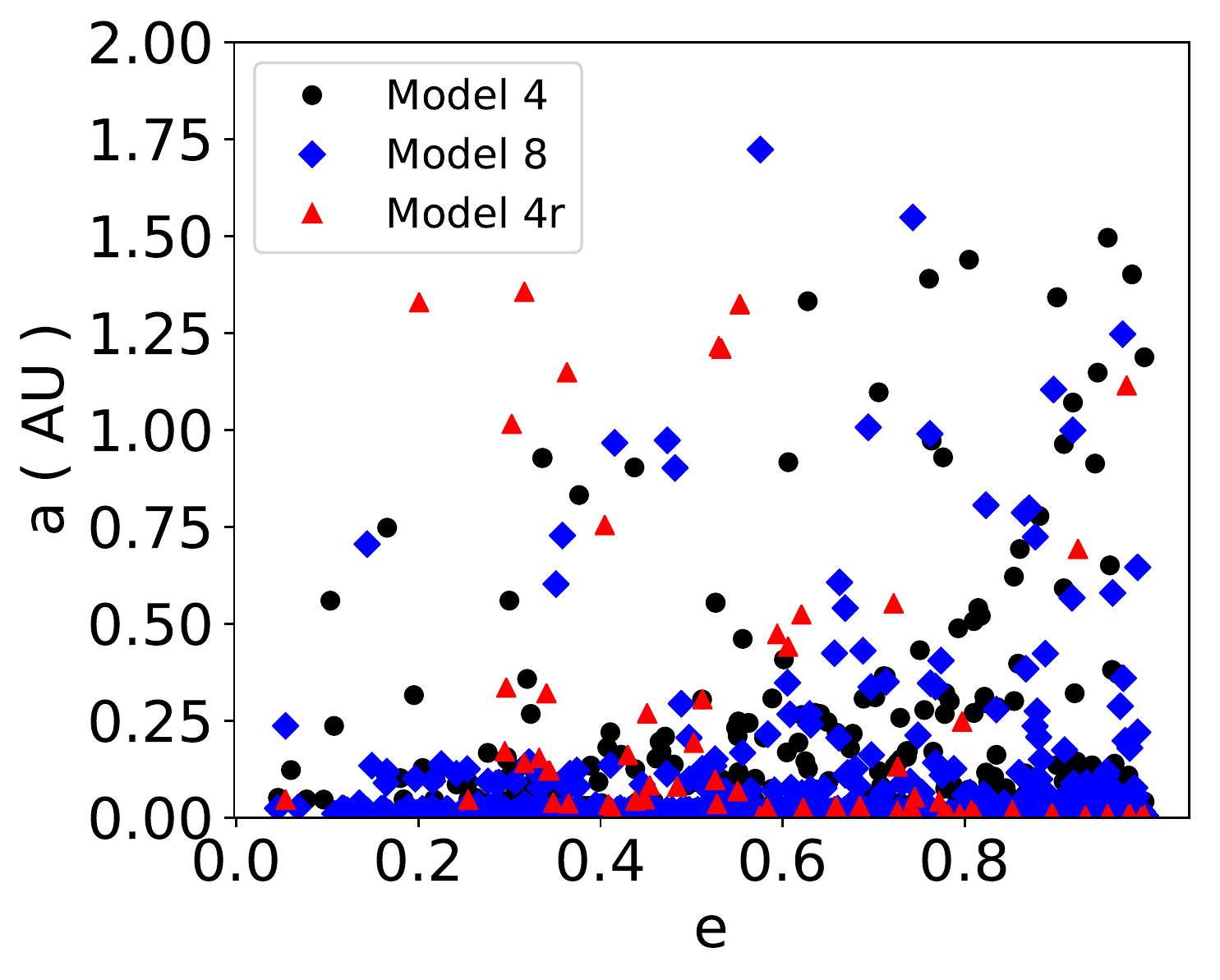}}
\end{minipage}
\caption{Orbital parameters (semi-major axis and eccentricity) for all HVBs produced in Model 1/1r/5 (top-left), Model 2/2r/6 (top-right), Model 3/3r/7 (bottom-left) and Model 4/4r/8 (bottom-right).}
\label{fig:hvb_sembin}
\end{figure*}

We performed $10^4$ simulations of close encounters for each combination of the parameters given in Table\,\ref{tab:model}, for a total of $6.8\times10^4$ scattering experiments. We uniform sample within the range given for the different parameters in Tab.\,\ref{tab:model}. We consider $8$ main models, where we run scattering events for a wide range of initial conditions, fixing all the parameters of the 2+2 quadruple star but one in order to study its effect on the relative channel probabilities and inferred rates. In Model 1, we study the fate of quadruples as a function of $\ain=a_{in,1}=a_{in,2}$, the semi-major axis of the two inner binaries. In Model 2, we examine the outcomes of quadruples as a function of the semi-major axis of one of the inner binary stars $a_{in,1}$. In Model 3, quadruples have different outer semi-major axis, while, in Model 4, we study the role of the stars mass. In Models 5/6/7/8 we consider the same initial conditions for quadruples as in Models 1/2/3/4, but half the values for $a_{in,1}$, $a_{in,2}$ and $\aout$. Finally, we perform additional runs of Models 1/2/3/4 where we consider finite stellar radii (Models 1r/2r/3r/4r). Fig.\,\ref{fig:scattex} illustrates an example of scattering for Model 1 (with $\ain=a_{in,1}=a_{in,2}=0.05$ AU), resulting in one HVS, one S-star and a binary S-star. Note that all stars left bound to the MBH have large eccentricities \citep{frs18}.

We report the branching ratios (BRs), i.e. the probabilities of different outcomes, in Fig.\,\ref{fig:br_all} for Models 1/2/3/4 (with the relative Poisson error bars), as a function of $\ain=a_{in,1}=a_{in,2}$, $a_{in,1}$, $\aout$ and $m_*$. As discussed in Sect. \ref{sect:meth}, we report the relative BRs of the $5$ main outcome classes based on the multiplicity of stars after the close interaction with the MBH, without distinguishing among cases in which the stars remain bound to the MBH as S-objects or escape as hypervelocity objects. In Model 1, the probability for the outcome 4S is an increasing function of $\ain$, while the probability for the channel 2B decreases for larger inner semi-major axis and the outcome 1B-2S is roughly independent of $\ain$. We found a similar trend in Model 2 as function of $a_{in,1}$, apart from the BR for 1B-2S channel that increases with larger semi-major axis of the first inner binary. In Model 3, the outcome probability of channel 1B-2S is nearly constant with $\aout$, while the probability for 4S decreases and for 2B increases. Finally, all the BRs are nearly independent of the stellar mass. In all the models, the probabilities of the channels 1T-1S and 1Q are negligible. We find similar trends for Models 5/6/7/8. We show in Table\,\ref{tab:br1} the BRs for Model 1 for all the possible outcomes, where we distinguish between bound and hypervelocity objects, as function of $\ain=a_{in,1}=a_{in,2}$. The probability of producing HVBs is very small, $\lesssim 2-4\%$ in all the cases, but we found nearly two-three times the number of HVBs originated in the triple disruption scenario \citep{fgu18}. When finite stellar radii are taken into account, mergers occur in about $18-70\%$ of the encounters, but still the relative BRs of the different channels have always the same qualitative behaviour, but smaller with respect to the point mass cases. We found the largest fraction of collisions in Model 1 with $\ain=0.05$ AU ($\approx 70\%$), which decreases as the inner binaries become wider.

\citet{fgu18} studied the triple disruption scenario and interpreted the low BR for HVBs by means of the typical involved energies \citep{yut03}. The same considerations hold in the case of a 2+2 quadruple star, where a typical scattering would lead to the disruption of the quadruple into two binaries (according to the initial conditions described in Sect. \ref{sect:meth}). The typical increase in the specific energy $\delta E$, given by the former binding energy of the quadruple system, is much larger than the binding energy of each of the two binaries, which can be converted into internal energy of the binary stars themselves, which become wider. In this case, the binaries may be tidally disrupted by the MBH, which does not necessarily lead to the production of an HVS since the HVS ejection velocity $v_{ej}\propto a^{-1/2}$ \citep{hills88,brm06}. In the case the widening of the semi-major axis is not sufficient for the tidal disruption, the former inner binaries can either remain on a bound orbit around the MBH (binary S-stars) or be ejected as HVBs. We found that 2+2 quadruples produce $\approx 2-3$ times the number of HVBs originated as a consequence of a triple star disruption. Since in this case there is no difference between the lighter single mass and the more massive binary as in the case of a triple disruption \citep{fgu18} and a quadruple star has a larger reservoir of energy, we expect that the probability of producing an HVB is larger than in the case of the triple disruption.

\begin{figure*} 
\centering
\begin{minipage}{20.5cm}
\subfloat{\includegraphics[scale=0.58]{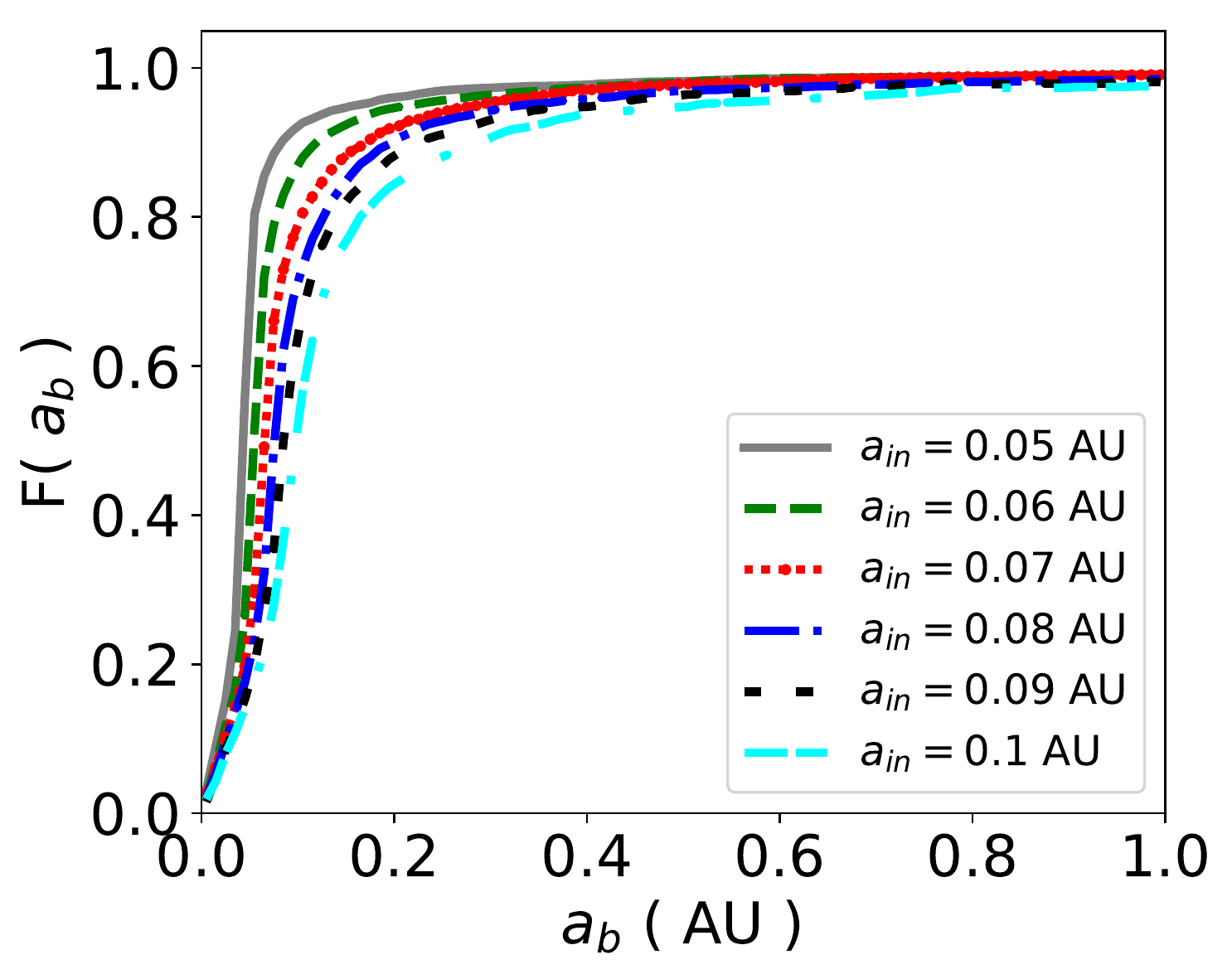}}
\subfloat{\includegraphics[scale=0.58]{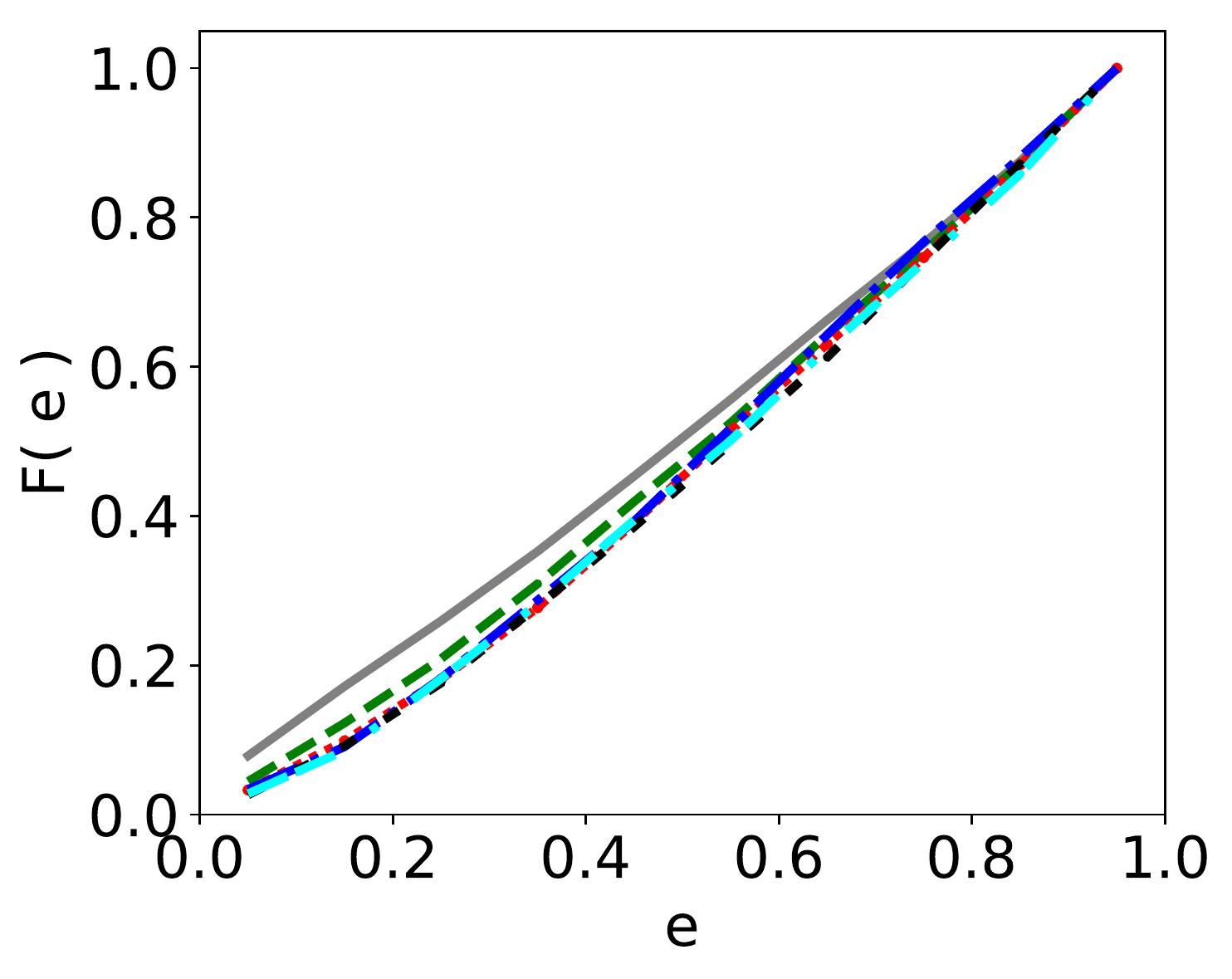}}
\end{minipage}
\caption{Cumulative distribution function of binary S-stars semi-major axis (left) and eccentricity (right) as function of $\ain$ for Model 1. More than $\approx 90$\% of binary S-stars have $a_b\lesssim 0.5$ AU to avoid tidal disruption by the MBH. The cumulative distribution of eccentricities is nearly independent on the initial $\ain$ and is $\propto e$, i.e. the eccentricity distribution is constant.}
\label{fig:binbh}
\end{figure*}

In Fig.\,\ref{fig:vel_sin} and Fig.\,\ref{fig:vel_bin}, we illustrate the velocity distribution for HVSs and HVBs in the different models, when all the simulations with different $\ain$, $a_{in,1}$, $\aout$ and $m_*$ are considered, respectively. The distribution for HVSs is peaked around $2000\kms$ in all models, with a tail extending up to $6000\kms$. For Models 5/6/7/8, the distribution has a larger fraction of stars with velocities $\gtrsim 2000\kms$ as a consequence of the smaller initial $a_{in,2}$, $a_{in,1}$ and $\aout$. The velocities of the HVBs produced in the simulations similarly have a peak at $\approx 1400\kms$, with outliers up to $\approx 4000\kms$. The smaller peak in the distribution of HVB ejection velocities can be explained in terms of the different masses of the escaping objects. In our simulations, the stars in the quadruple have all the same mass, thus HVBs are twice more massive than HVSs. When different masses are involved, the escape velocity of the ejected object scales as \citep{brm06}
\begin{equation}
v_{esc}\propto \left(\frac{m_{esc}}{m_{tot}}\right)^{1/2}\ ,
\end{equation}
where $m_{esc}$ is the mass of the escaper and $m_{tot}=m_{esc}+m_{bound}$ is the total mass of the escaper and of the object that remains bound to the MBH. In our simulations
\begin{equation}
v_{HVS}=\sqrt{2}v_{HVB}\ ,
\end{equation}
which accounts for the ratio of the peak velocities ($\approx 2000/1400\approx 1.4$) of HVS and HVB distributions.

Figure \ref{fig:hvb_sembin} shows the semi-major axes and eccentricities of the HVBs produced as a consequence of the 2+2 quadruple disruption. Most of the HVBs have small semi-major axis ($\lesssim 1.0$ AU) and large eccentricity ($\gtrsim 0.3$), as expected from the previous theoretical considerations. We note that Models 5/6/7/8 produce about three times more HVBs than Models 1/2/3/4 as a consequence of the smaller initial inner and outer semi-major axis. The tighter the inner binaries, the larger the energy reservoir that can be exchanged during the five-body encounter and the larger the probability of generating HVBs.  

\citet{gou03} proposed that the innermost observed S-stars in the GC could be the former companions of ejected HVSs. Recently, \citet{lii18} and \citet{nao18} have discussed the role of binary stars in shaping the properties of the disk of stars near the GC, but still it is challenging to resolve the binary sources and distinguish them from single stars. Triple and quadruple star disruptions naturally bring binaries in the vicinity of the MBH. Figure \ref{fig:binbh} shows the cumulative distribution function of binary S-stars semi-major axis and eccentricity as a function of $\ain$ for Model 1, which shows a clear correlation between the final and initial binary semi-major axis. More than $\approx 90\%$ of binary S-stars have $a_b\lesssim 0.5$ AU to avoid tidal disruption by the MBH and the eccentricity cumulative distribution is nearly independent of the initial $\ain$ and is $\propto e$, i.e. the eccentricity distribution is constant. Future data, as those expected by the James Webb Space Telescope, may observe near-infrared sources with higher photometric precisions, hence helping in identifying binaries in the GC.

\section{Discussions and Conclusions}

Most of the observed HVSs support the standard Hills scenario, but the $\sim 9\msun$ main-sequence HVS HE0437-543 and the candidate HVB require different production mechanisms than the standard binary disruption. While HE0437-5439 is likely to come from the centre of the Large Magellanic Cloud \citep{erkd18}, the candidate HVB as well as the other HVSs whose travel times exceed their main sequence lifetime would require disruption of triple stars by the MBH, which would produce a HVB that can later evolve into a blue straggler star as a consequence of binary evolution \citep{per09}. However, the recent detailed N-body simulations by \citet{fgu18} have showed that the ejection rate is too low to explain these hypervelocity outliers.

In this work, we discussed the 2+2 quadruple tidal disruption scenario by means of high-accuracy scattering experiments. While less abundant than triple stars, 2+2 quadruple stars constitute $\approx 4\%$ of the stars in the solar neighborhood \citep{tok14a,tok14b} and have been shown to have interesting dynamics dealing with white dwarf-white dwarf collisions \citep{fan17} and gravitational wave emission \citep{set18}. We studied a wide range of parameters of the quadruple stars and performed simulations with both point masses and finite stellar radii. We found that only a very small fraction ($\lesssim 2-4\%$) of encounters results in the ejection of a HVB. Taking into account that the estimated ejection rate for the Hills mechanism is $\approx 10^{-6}-10^{-4}\yr^{-1}$ \citep{yut03} and that 2+2 quadruple stars are roughly a third of triple stars, we conclude that also 2+2 quadruple disruption leads to a HVB production rate of $\approx 1\gyr^{-1}$, similarly to the triple disruption scenario. We conclude that quadruple disruption, as long as triple disruption, is an unlikely source of HVBs, which would require a non Galactic-Centre origin, as the origin from external galaxies \citep{bou2017}, from the Galactic disk \citep{per12} or as a consequence of the disruption of a cluster \citep{fck17} or dwarf galaxy \citep{abadi09}. Proper motions from the \textit{Gaia} mission should constrain the HVSs origin in the near future.

\section{Acknowledgements}

GF thanks Seppo Mikkola for helpful discussions on the use of the code \textsc{archain}. Simulations were run on the \textit{Astric} cluster at the Hebrew University of Jerusalem. This research was partially supported by an ISF and an iCore grant.

\bibliographystyle{mn2e}
\bibliography{biblio}
\end{document}